# Frequency-Modulated Magneto-Acoustic Detection and Imaging: Challenges, Experimental Procedures, and B-Scan Images

Miaad S. Aliroteh, *Student Member, IEEE*, Greig C. Scott, Member*, IEEE*, and Amin Arbabian, *Member, IEEE*

*Abstract*— Magneto-acoustic tomography combines near-field radio-frequency (RF) and ultrasound with the aim of creating a safe, high resolution, high contrast hybrid imaging technique. We present continuous-wave magneto-acoustic imaging techniques, which improve SNR and/or reduce the required peak-to-average excitation power ratio, to make further integration and larger fields of view feasible. This method relies on the coherency between RF excitation and the resulting ultrasound generated through Lorentz force interactions, which was confirmed by our previous work. We provide detailed methodology, clarify the details of experiments, and explain how the presence of magneto-acoustic phenomenon was verified. An example magneto-acoustic B-scan image is acquired in order to illustrate the capability of magneto-acoustic tomography in highlighting boundaries where electrical conductivity alters, such as between different tissues.

## I. INTRODUCTION

IN this work, we investigate a multi-modal imaging technique – Magneto-Acoustic Imaging (MAI) – that we believe can evolve into a scalable, economical, portable, and non-hazardous imaging system. Magneto-acoustic imaging (MAI) is a hybrid method combining ultrasound (US), for high spatial resolution, and near-field low-frequency RF for deep penetration and tissue electrical conductivity contrast [1-12]. This phenomenon was first introduced in [1] where it was shown that current-carrying media, in the presence of static or alternating magnetic fields, result in Lorentz forces that generate detectable acoustic vibrations. Overtime, MAI was further refined with major contributions from [2-12]. MAI has no hazardous radiation, unlike CT and PET, and generates tissue contrast from dielectric properties. For example, cell membrane structure as well as macroscopic structures including vascularization, angiogenesis in cancers, and necrotic cores will all influence electrical conductivity. MA signals increase linearly with magnetic and electric field strength. MRI techniques in the form of RF Current Density Imaging (RF-CDI) and Magnetic Resonance Electrical Impedance Tomography (MR-EIT) can also capture conductivity contrast [13-15]. However, MAI can tolerate substantially higher field non-uniformity than MRI, which leads to the possibility of integration into smaller and portable form factors.

MAI has a well-established history although it has yet to obtain widespread adoption due to challenges in scaling up to the human body. An early exploration of magneto-acoustic phenomena was done by [1], which non-invasively quantified the magnitude of 3kHz alternating currents, as low as 7μA (limited only by amplifier noise), in the hamster abdomen with a 0.2T static magnetic field. Despite their results, they raised concern about successfully scaling MAI to the human body. In another pioneering work in MAI [2], the MAI was performed on a block of bacon consisting of multiple layers of muscle and fat. This was achieved with a 4T magnetic field and a 500V pulse excitation corresponding to 1.25kW peak power in a 50Ω coaxial transmission line. It was estimated that pressure levels below 1Pa are produced with magnetic fields less than 1T based on the safe levels of electromagnetic excitation and nerve-stimulation thresholds. Later work by [3] extended MAI by introducing non-contact, inductive excitation through the induction of eddy currents within the target being imaged. Here, it was estimated that with 1T magnet field, a pressure level of 15mPa would result from a 200A/m$^2$ current density in an object with 0.2S/m conductivity, corresponding to 1000V/m induced electric fields. In experiments, a permanent magnet was used to obtain 0.1T magnetic field within the sample. With an excitation scheme inducing 25V/m electric fields, [3] obtained measurements with an SNR between 6dB to 10dB after 100 averages and produced an image from a metal wire loop within the sample. The work in [3] was later extended in [4] where simple 3D saline gel phantoms as well as multilayered muscle-fat tissue were successfully imaged using non-contact induction of pulsed eddy currents. More recent work by [5-12] introduced advanced image reconstruction techniques to further improve spatial resolution and reduce artifacts. Moreover, [6] further extended [4] by implementing advanced reconstruction techniques on experimental data in order to quantify the conductivity distribution of multiple targets within the field of view. Contrast agents based on antibody-conjugated magnetic

Manuscript received August 24, 2015. This work was supported in part by the U.S. Department of Defense Advanced Research Projects Agency (DARPA) as part of the Methods for Explosive Detection at Standoff (MEDS) program as well as National Institutes of Health (NIH) under grants R01EB008108 and P01CA159992.
Miaad S. Aliroteh is a Ph.D. candidate at the Electrical Engineering Department, Stanford University, Stanford, CA 94305 USA (miaad@stanford.edu).
Greig C. Scott is a Senior Research Engineer with the Magnetic Resonance Systems Research Laboratory (MRSRL), Stanford University, Stanford, CA 94305 USA (greig@mrsrl.stanford.edu).
Amin Arbabian is an Assistant Professor at the Electrical Engineering Department, Stanford University, Stanford, CA 94305 USA where he is also a School of Engineering Frederick E. Terman Fellow (arbabian@stanford.edu).

nanoparticles were introduced in [7] for labeling particular tissues of interest. Furthermore, [7] also demonstrated higher resolution, down to cellular level, by using higher acoustic frequencies and bandwidths at the cost of small field of view and lower penetration depth.

Over the decade, MAI has seen major improvements in implementation, resolution, and reconstruction algorithms. One area that requires further refinement, before MAI can achieve widespread adoption, is the peak power reduction and optimization at the system level. In fact, [2] illustrated that even at high magnetic flux densities of 4T, high peak powers levels, greater than 1kW, are still necessary for the excitation mechanism. In addition, [4-6] used a permanent magnet with reduced static magnetic flux density, about 0.1T, and consequently required even higher powers for their excitation. The 1μs pulsed inductive excitation schemes in [4-6] generated between 0.01T to 0.1T of magnetic field within an approximately 125cm³ volume corresponding to a required peak power greater than 10kW. These peak power levels would have to be further increased in order to scale to the human body and indeed this is not a trivial task [1]. Even if possible, such a MAI system is likely to be bulky in size and costly. Although this may be acceptable in a clinical setting, low cost and portable applications such emergency imaging of hemorrhages, stroke damage, and other paramedic scenarios cannot readily use such a system.

In this work we perform an in-depth study of a technique for peak power reduction that builds on previous work in MAI [16]. MAI can be performed with coherent processing techniques that rely on amplitude and phase, namely, (Step) Frequency Modulated Continuous Wave (FMCW/SFCW) excitation [16]. With SFCW RF excitation, the peak power requirement is reduced by a factor of 4000 over pulsed excitation with equivalent resolution, contrast, and SNR. Thus, in this way, the CW approach makes MAI truly a more practical and economical alternative for low-cost imaging based on conductivity contrast. It is important to note we are not proposing replacing previous MAI implementations with this new technique but rather encouraging the addition of this technique to existing MAI implementations. The main focus of this paper is to: (i) provide extensive CW MAI theory, (ii) provide detailed methodology, (iii) explain the experiments performed in detail and how we verified detection of actual MA signals, and (iv) demonstrate an example B-scan image.

## II. Theory

### A. Working Principle

The theory for the pulsed-based magneto-acoustic imaging is explained by [2, 3] and others. In this work we will provide theory for the continuous-wave scenario. RF excitation impresses currents in tissue that interact with a static DC magnetic field to generate Lorentz forces. The RF excitation is in the same frequency range as US. US pressure waves are generated at boundaries between tissues of different conductivity where there is a gradient of conductivity [3]. This derivation assumes that the time rate of change in the magnetic field ($B_1$) is either negligible ($\nabla \times E \approx 0$), as it is for near-field, or perpendicular to $B_0$, as in the case of eddy current induction. This assumption is not valid for [4 – 6] where $B_1$ and $B_0$ are parallel, and where $\sigma \nabla \times E$ may dominate.

$$\nabla^2 p - \frac{1}{v^2}\frac{\partial^2 p}{\partial t^2} = \nabla \cdot (\boldsymbol{J} \times \boldsymbol{B_0}) \tag{1}$$

$$\nabla \cdot (\boldsymbol{J} \times \boldsymbol{B_0}) = \boldsymbol{B_0} \cdot (\nabla \sigma \times \boldsymbol{E} + \sigma \nabla \times \boldsymbol{E}) \tag{2}$$

$$\frac{1}{v^2}\frac{\partial^2 p}{\partial t^2} - \nabla^2 p \approx -\boldsymbol{B_0} \cdot (\nabla \sigma \times \boldsymbol{E}) \tag{3}$$

MA excitation has been proposed with non-contact capacitive electrodes, non-contact induction coils, or directly with current-injecting electrodes. This present work uses contact, injecting electrodes for a proof-of-concept design, however continuous-wave techniques can be implemented with all of these excitation mechanisms. It would seem non-contact eddy-current induction is an ideal excitation mechanism. Nevertheless, this approach requires delicate engineering to work reliably. More importantly, induction only creates mirror-image currents of the source and cannot create deep electric fields. Contact electrodes do not have this limitation in cases where their deployment is possible.

### B. Harmonic MA Theory

Consider Fig. 1 where the tissue sample is surrounded by water. The entire space is divided into small enough cubes such that all physical quantities are uniform in each cube. For simplicity, posit that the static DC magnetic field ($B_0$) is aligned with the x-axis, the current density within tissue aligned with the y-axis, and so the expected Lorentz force will align along the z-axis. Considering one small cube of tissue, the Lorentz force will cause it to displace from its nominal position with an acceleration and velocity.

Now suppose that a reasonable estimate for the expected MA pressure level is desired. By assuming *J* and other physical quantities are sufficiently uniform and similar in all the cubes, this problem can be treated as a one dimensional problem in the z-axis; with the conductivity boundary at the $z=0$ plane. This is solved using the 1D non-homogeneous wave equation in (3) by assuming continuous-wave operation (phasor notation) and that the conductivity varies only in the z-direction. This results in (4) which is turned into the non-homogenous Helmholtz equation (5-6):

$$\frac{1}{v^2}\frac{\partial^2 p}{\partial t^2} - \frac{\partial^2 p}{\partial z^2} = -B_0\hat{x}\cdot\left(\frac{\partial \sigma}{\partial z}\hat{z}\times E_y\hat{y}\right) = \frac{\partial \sigma}{\partial z}B_0 E_0 e^{j\omega t}$$
$$= C(z)e^{j\omega t} \quad (4)$$

$$p(z,t) = A(z)e^{j\omega t} \quad (5)$$

$$(\nabla^2 + k^2)A(z) = \delta(z) * C(z) \quad (6)$$

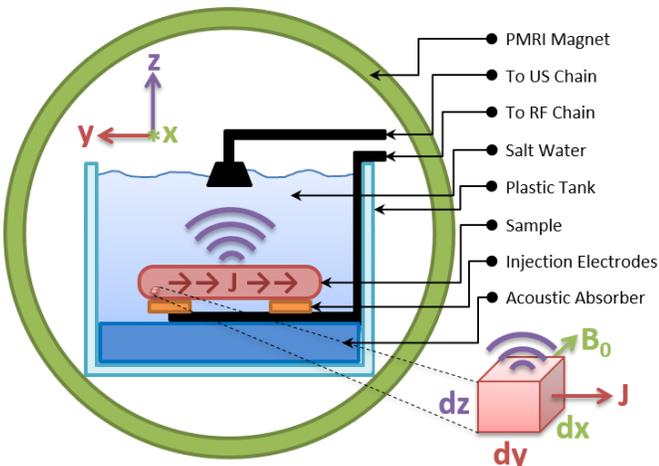

**Fig. 1**. Experimental setup for detecting MA signals from tissue using rigid-micro-coax-driven current-injecting electrodes. The excitation currents interact with the static magnetic field producing detectable ultrasound vibrations [16].

Note that the non-homogenous term in (4) is written in explicit phasor notation, $C$, with a complex exponential in time domain whereas the Helmholtz equation in (6) is written in phasor notation in $\omega$-frequency domain with the non-homogenous term as the convolution of $C$ and the Dirac function $\delta$. In a 1D problem, the solution to the non-homogenous Helmholtz equation with $\delta$ as the non-homogenous term has a simple closed form. Thus the solution of (6), $A$ will be the convolution of that solution and $C$:

$$A(z) = \frac{e^{-\frac{j\omega}{v}|z|}}{2\cdot j\omega/v} * C(z) \quad (7)$$

Additionally, the gradient of conductivity resolves into a positive and negative Dirac function $\delta$ corresponding to each boundary of the sample in Fig. 1:

$$\frac{\partial \sigma}{\partial z} = \Delta\sigma\big(\delta(z) - \delta(z-\Delta z)\big) \quad (8)$$

where $\Delta z$ is defined as the finite thickness of the target tissue slab with assumed uniform conductivity. Thus (7) is further simplified to its final form:

$$A(z) = \frac{\Delta\sigma E_0 B_0}{2\frac{j\omega}{v}}\left(e^{-\frac{j\omega}{v}|z|} - e^{-\frac{j\omega}{v}|z-\Delta z|}\right) \quad (9)$$

If (9) was further simplified by assuming a thin membrane of half an acoustic wavelength thick ($\Delta z = \pi\cdot v/\omega$) with $k=\omega/v$ as the wave number then finally the pressure would be:

$$p(z,t) = \frac{\Delta\sigma E_0 B_0}{j\omega/v} e^{j\left(\omega t - \frac{\omega}{v}|z|\right)} \quad (10)$$

In practice the solution in (10) serves as a good order-of-magnitude hand-calculation for the general 3D problem where diffraction becomes an issue. Nevertheless, the procedure in (6-10) can also be implemented for the 3D problem. Hence, accurate yet simple order of magnitude calculations (where diffraction and other acoustic issues are neglected) can be performed with (9) or (10). Such hand calculations for experimental setups similar to Fig. 1, described in detail later, suggest an expected pressure of 18mPa$_{peak}$ compared to hydrophone-calibrated measurements that demonstrate pressures in the 10mPa$_{peak}$ to 40mPa$_{peak}$ range (depending on exact conductivity change) using less than 1W of power or roughly 580 A/m$^2$ of change in equivalent current density in a 0.13T magnetic field. For a typical fat-muscle interface the pressure levels would be approximately an order of magnitude larger due to greater conductivity changes. Further, considering the limitations of permissible specific absorption rates, SAR = 5W/kg, 0.2T magnetic field, and a conductivity changes of 0.5S/m between fat and muscle, the permissible current density decreases to 50 A/m$^2$. This would result in an expected pressure of only 2mPa$_{peak}$. The mechanical noise of an ideal 1.3cm$^2$ transducer is about $14\,\mu Pa/\sqrt{Hz}$ which translates to 0.07mPa RMS noise level with averaging to an equivalent noise bandwidth of 25Hz [17, 18]. In order to detect these lower pressure levels at safe average SAR levels, averaging as well as long-term duty cycling may be necessary. In addition, optimized transducers, such as Capacitive Micro-machined Ultrasound Transducers (CMUTs), and custom receiver electronics may be advantageous [19].

### C. FMCW & SFCW MA Theory

FMCW/SFCW are continuous wave (CW) techniques that reduce the peak power requirements of an imaging system while maintaining the same average power, SNR, averaging time, etc. This technique was first employed in radar [20, 21]. In FMCW, shown in Fig. 2, a linear frequency modulated (LFM) RF chirp signal is generated, and amplified yielding the transmit (Tx) signal at the target region:

$$Tx = A\cdot cos\left(2\pi\left(f_0 + \frac{\Delta f}{T}t\right)\right) \quad (11)$$

Here, $A$ is the transmit amplitude, $f_0$ is the minimum excitation frequency, $T$ is the modulation period, and $\Delta f$ is the modulation bandwidth. In the presence of a static magnetic field ($B_0$) a coherent acoustic signal is generated by differential Lorentz forces, detected by the transducer, and amplified by the receive chain (Rx). The Rx signal frequency lags behind the instantaneous Tx frequency commensurate with the range of the target and the speed of sound as in (12):

$$Rx = B \cdot cos\left(2\pi\left(f_0 + \frac{\Delta f}{T}(t-\tau)\right)\right) \quad (12)$$

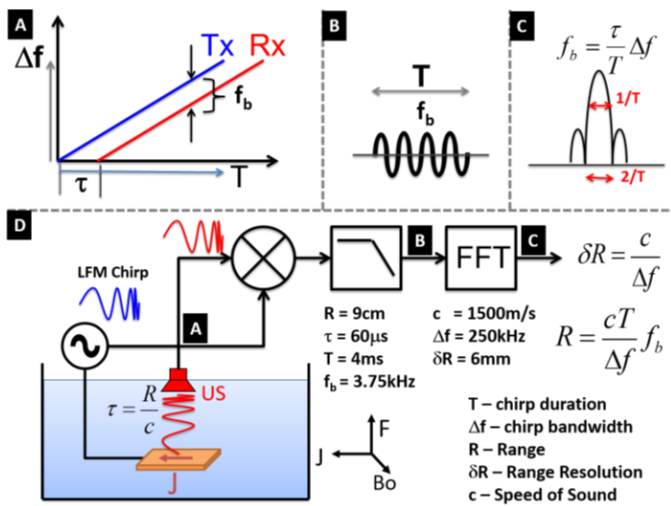

**Fig. 2**. The continuous-wave imaging technique using linear-frequency-modulated (LFM) chirp signals. (A) LFM chirp signal. (B) Demodulated signal in time. (C) Demodulated signal in frequency. (D) FMCW block diagram [16].

Here, $B$ is the receive amplitude and $\tau$ is the acoustic delay. The original LFM signal is used to demodulate, by complex multiplication (13) and low-pass filtering (14), the Rx signal into a sinusoid of frequency $f_b$:

$$Tx \cdot Rx = \frac{AB}{2}\left[cos\left(2\pi\frac{\Delta f}{T}\tau\right) + cos\left(2\pi\left(2f_0 + \frac{2\Delta f}{T}t - \frac{\Delta f}{T}\tau\right)\right)\right] \quad (13)$$

$$LPF\{Tx \cdot Rx\} \propto cos\left(2\pi\frac{\Delta f}{T}\tau\right) \rightarrow \frac{\Delta f}{T}\tau = f_b, \ \tau = \frac{R}{c} \rightarrow R = \frac{cT}{\Delta f}f_b \quad (14)$$

Frequency $f_b$ is directly proportional to the lag time and hence the target range. This can be viewed as a form of cross correlation between the transmitted and received signals. In practice, data are first apodized by a Hanning window and Fourier transformed.

Similar to radar, the range resolution of FMCW is dependent on the modulation bandwidth and the linearity of the frequency modulation while its final SNR depends on the fidelity of the modulation [20, 21]. Due to limitations in our present instrumentation, it can only synthesize a frequency chirp as a continuous series of coarse quadratic phase steps every 2μs, leading to spurious errors that increase the apparent spectral noise.

In SFCW, we perform a similar process but with a "staircase" of $N$ discrete frequency steps that will ultimately phase encode range. The $n^{th}$ step is:

$$Tx(n) = A \cdot cos\left(2\pi\left(f_0 + \frac{\Delta f}{N-1}nt\right)t\right) \quad (15)$$

SFCW requires the signal to reach a steady state frequency response whereas the FMCW technique requires the signals to be continuously in a transient state. The steady state magneto-acoustic signal is a superposition of delayed sinusoids from potentially multiple sources as in Fig. 3, and (16):

$$Rx(n) = \sum_{k=0}^{N-1} a_k \cdot cos\left(2\pi\left(f_0 + \frac{\Delta f}{N-1}n(t-\tau_k)\right)t\right) \quad (16)$$

Upon demodulation, discrete signal sources of time delay $\tau_k$ yield stepped phase increments:

$$\varphi_k(n) = 2\pi\left(f_0 + \frac{\Delta f}{N-1}\right)n\tau_k \quad (17)$$

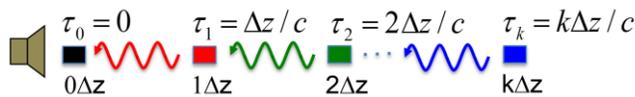

**Fig. 3**. In the SFCW technique, the sources reach steady state and produce acoustic signals that mirror the sinusoidal excitation. The time delay is encoded as a phase delay at the receiver, as in (17), and later decoded during reconstruction.

$$\tau_k = \frac{k\Delta R}{c}, \max(\varphi_k(n)) = 2\pi \rightarrow R_{max} = \frac{c}{\Delta f}, \ \Delta R = \frac{R_{max}}{N} \quad (18)$$

where $N$ is the number of frequency samples, $\varphi_k$ is the encoded phase from the $k$th target, $\Delta R$ is the range resolution, and $R_{max}$ is maximum detectable range before aliasing. Here, the sequence of demodulated complex weights represents samples of the frequency response. They are first Hanning windowed, and then applied to a discrete Fourier transform to recover range. SFCW does not have strict linearity and fidelity requirements but it will have limited detection range above which the range will alias due to the inherent frequency-domain sampling [20] as in (18). SFCW is analogous to a network analyzer time domain mode which also samples a spectrum, hence leading to potential aliasing, as well as how it filters noise with an FFT operation, thus increasing SNR for coherent signals through process gain. In comparison FMCW has no aliasing limitation although it requires sufficient modulation fidelity. Nonetheless, SFCW may be more practical to implement since it is amenable to digital and hence automatic tuning. Finally, both SFCW and FMCW have theoretically identical SNR efficiency when implemented correctly.

### III. EXPERIMENTAL SETUP

Experiments are carried out in a bidirectional pulsed electromagnet with field strengths of ±0.13T similar to the system in [22, 23]. The waveform generation of the RF excitation, US signal acquisition, and post processing are performed with the MEDUSA acquisition system [24] and Matlab. A 200W RF peak power amplifier is transformer-coupled – to mitigate leakage and EMI – to current-injecting, strip electrodes such that the RF power delivered to the tissue sample is between $1W_{peak}$ to $10W_{peak}$. A 20mm x 20mm and 6mm thick slice of chicken breast is placed on top of the injection electrodes as in Fig. 1. Note that the electrodes are aligned parallel to the static magnetic field ($B_0$) so that they would not generate MA signals. Only the current flowing within the tissue sample (and nearby salt water) produced MA signals. Copper foil acts as a better controlled source of MA signals than biological tissues as observed in experiments. As shown in Fig. 1, 2, and 4, a hydrophone-calibrated immersion transducer (1MHz, V303 from Olympus) is positioned 4cm to 9cm above the tissue in a tank filled with salt water (about 0.6 S/m) and with acoustic absorbers (Precision Acoustics, Aptflex F28) lining the tank bottom. The transducer is directly connected, with no matching network, to a 38dB LNA (AD797) with less than 2nV/√Hz of input referred noise. The LNA, which is battery powered and hence isolated, is capacitor-transformer coupled – to mitigate leakage and EMI – to a commercial 30 dB amplifier (Parametric 5055PR) whose output connected to the acquisition system and an oscilloscope. The transformer couplings, on both transmit and receive, reduces common-mode leakage from the transmit chain into the receive chain. Total leakage is -100dBc and -130dBc for the capacitive electrode and current-injecting electrode scenarios, respectively.

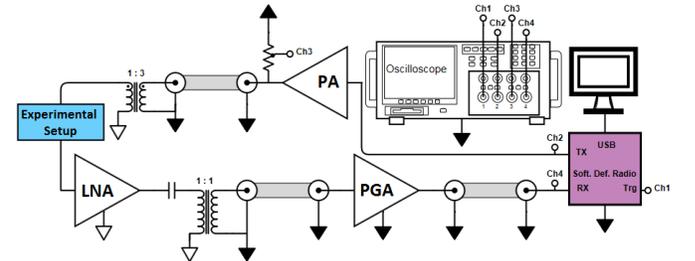

**Fig. 4**. (A) Experimental schematic: instrumentation is isolated from the electrodes, transducer, and LNA. The experimental setup is depicted in detail in Fig. 1.

## IV. RESULTS AND DISCUSSION

### A. Initial Challenges

In this work, several challenges are overcome in order to have certainty in the presence or lack of MA signals during the experiments. First, we observed that transducers (A314) with matching circuits (ferrite-core inductors) will become ineffective inside the static magnetic field due to core saturation as shown in Fig. 5. Thus, an unmatched transducer (V303) has to be used with a short connection to a low noise amplifier (LNA) whose output is then compatible with the rest of the 50Ω system. This arrangement with the V303 transducer shows no noticeable degradation within the magnetic field.

Second, RF excitation leakage can potentially be large enough to excite the transducer into producing ultrasound that is later reflected from the tank bottom and received, irrespective of the magnetic field. This is termed as parasitic echo generation as shown in Fig. 6 where the magnetic field is off. This phenomenon can potentially appear disguised as a phantom MA signal if it is not attenuated (for CW RF excitation) or separated in time (for pulsed RF excitation and CW RF excitation with simple targets) from actual MA signals. To implement this attenuation experimentally, capacitive electrodes lining the tank walls were replaced with current-injecting electrodes floating at the tank bottom. Rigid micro-coax cable connects the injection electrodes to the rest of the transmit chain. Acoustic absorbers (Precision Acoustics, Aptflex F28) are also placed at the tank bottom to mitigate secondary reflections.

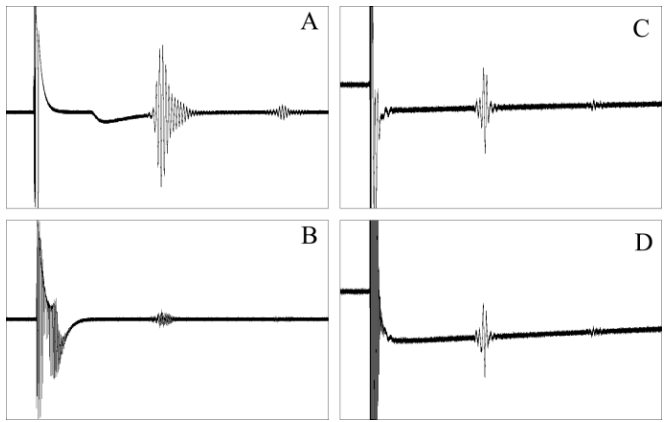

**Fig. 5**. Left: Pulse-echo signal levels from impedance-matched transducers (A314) with the static magnetic field off (A) were stronger than those with the magnetic field on (B). Right: Pulse-echo signal levels from non-matched transducers (V303) with the static magnetic field off (C) were nearly the same as those with the magnetic field on (D). The impedance matching is hampered by the static magnetic field as it saturates ferrite-core inductors in the matching circuit. Note: a secondary weaker echo is also present as expected.

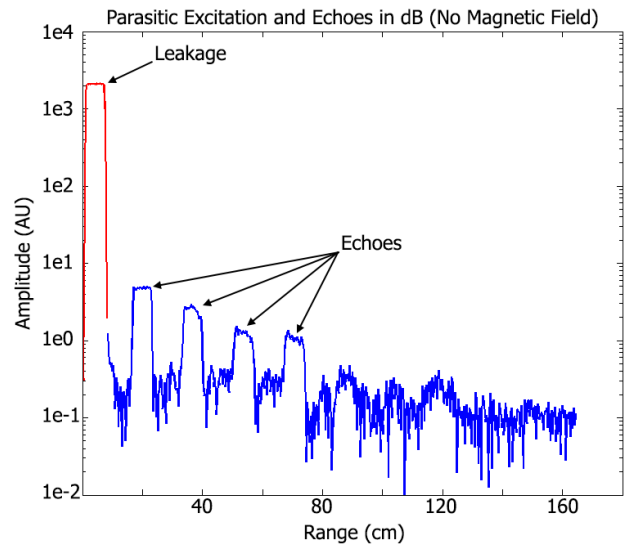

**Fig. 6**. RF leakage into the ultrasound transducer causes its excitation, launching acoustic waves which are later detected as echoes.

In actual MA imaging applications, such parasitic excitation of the receiver transducer will effectively produce ultrasound transmissions whose reflections will be detected and result in a conventional US image, not a MA image [25]. For pulsed MA imaging, this may possibly be avoided by temporarily, electrically shorting the transducer terminals, with MOS switches, during the RF excitation time-window. Another approach is to electrically shield the transducer [25], such as with a copper mesh enclosure, while still allowing good acoustic coupling. Finally, for CW MA imaging, a custom, intelligent transducer design as well as system design and layout are critical to reducing such parasitic US signals below the MA signal levels. For simple targets, a separation in time between MA signals and parasitic US signals also exists for the CW MA technique after the demodulation stage. This is used to verify and separate the presence of MA signals from parasitic US signals in both the pulsed and CW excitation schemes.

The third challenge in MA detection and imaging arises from the leakage of RF excitation from the transmit path into the receive path through the coupling of RF electric fields into the transducer. In addition, leakage cross-talk also results from electromagnetic interference (EMI) between transmit circuitry and receive circuitry through power supply lines and electric field coupling to coax-cables. This phenomena is attenuated by the use of transformers and chokes as well as the use of the current-injecting electrodes mentioned earlier. The MA signal polarity changes with the polarity of the static magnetic field while all other signals, including leakage, are unaffected. This differential processing method can be used to further attenuate leakage and other undesirable signals leaving only the MA signal component and noise. This technique is only limited by the matching in the system (e.g. the magnetic field polarity must be reversed while maintaining almost the same magnitude) and any drift the amplifiers and other circuitry. In practice a 40dB reduction is possible without excessive effort, as illustrated in Fig. 7. Further enhancement requires dynamic calibration to suppress time varying and drift terms.

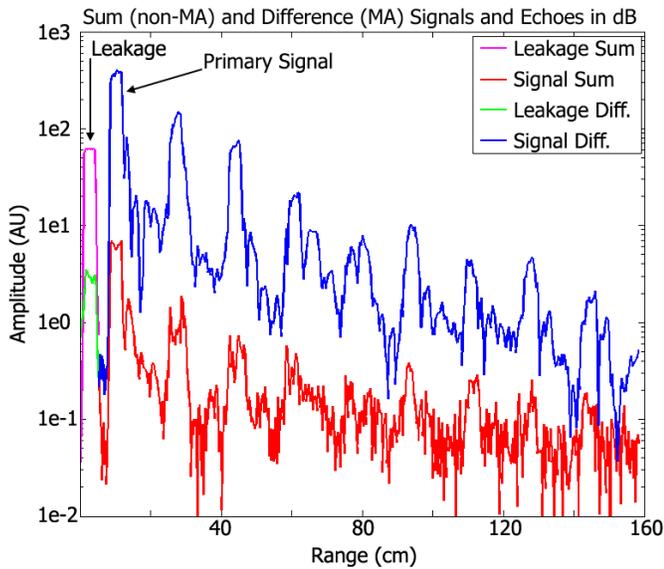

**Fig. 7.** MA signals, from a copper foil source, correlate with magnetic field reversal while leakage and non-MA signals do not. This allows for a "differential" MA signal detection with up to 40dB non-MA signal rejection.

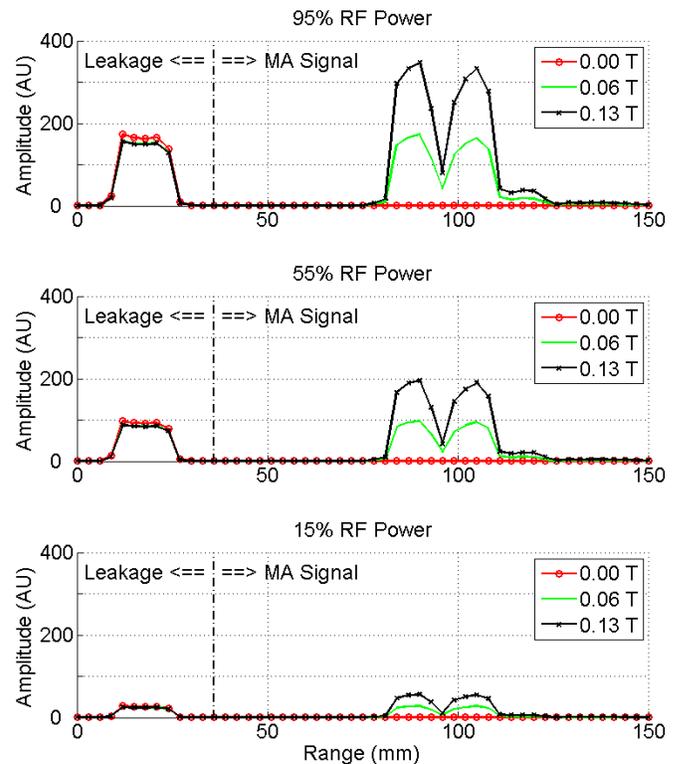

**Fig. 8**. Measured MA signals were proportional to the magnetic field strength and RF amplitude as expected. The shape of the acoustic waveform was altered due to reflected MA signals (which were slightly weaker as well) from the bottom of the container.

The fourth and final challenge is to isolate the MA effect from its cousins, the Thermo-Acoustic effect (TA) [26] and the Electroacoustic effect (EA) [27]. The TA effect is also produced using RF excitation, however here the US waves are produced as a result of thermal expansion. The TA effect is proportional to the power density and hence the square of excitation voltage [26] while it is independent of any magnetic fields. The EA effect, although not well understood, depends on the interaction of an electric field and ions at metal-electrolyte interfaces, and is linear with frequency and voltage. Both TA and EA effects are in contrast to the MA effect which is proportional to both the excitation voltage (electric field) and the magnetic field as observed in Fig. 8. Furthermore, in CW excitation, the acoustic frequency produced by the TA effect is twice the excitation frequency whereas in the MA effect, both frequencies are identical [26]. In pulsed excitation, the TA signal polarity is independent of the RF excitation polarity while the MA polarity changes with RF excitation polarity as shown in Fig. 7. In addition to magnetic field reversal, leakage attenuation, and parasitic US mitigation, these differences between MA and TA are exploited to verify that the detected signals are in fact MA signals and distinguishable from TA, EA, leakage, and parasitic US signals.

### B. MA Signals from Copper Foil Sample

Initial proof-of-principle experiments focused on the critical task of establishing the nature of received signals using a copper foil sample. Here, measurements with positive and negative magnetic fields are taken. The sum component reveals the non-MA signals that do not correlate with the magnetic field reversal including: leakage, parasitic echoes, and EA/TA effects. In contrast, the difference reveals only MA signals as only they are reversed when the magnetic field is reversed. In general, the accuracy of this technique is limited by the mismatch and drift in the system which was empirically determined to be about 1% as is illustrated in Fig. 9. Here, the MA signal originates 9cm away from the transducer and is echoed every 18cm as it reflects at the transducer-water interface and the tank bottom (no absorbers used). MA signals can also be discerned based on their linear dependence on the magnetic field strength and RF excitation level as opposed to other leakage or parasitic terms. In general however, once leakage is controlled through hardware techniques, no special signal processing techniques should be required to sift the MA signal from undesired interference. In another experiment, a copper foil target as well as acoustic absorbers, lining the tank bottom, are used. This time multiple measurements are made with the transducer displaced by 1.5mm horizontally (x-axis) at each step to produce the MA image show in Fig. 10.

### C. MA Signals from Tissue: Pulse vs. FMCW vs. SFCW

In order to draw a fair comparison between pulsed, FMCW, and SFCW approaches to MA detection, experiments using chicken breast tissue sample are performed. As such, acquisition and averaging time (4.2sec), receive chain amplification (68dB), resolution (6mm), and peak RF excitation power (10W) must remain the same across the test. Note that resolution limits the maximum pulse duration (and hence bandwidth) in the pulsed scenario and the bandwidth (250kHz) for FMCW and SFCW. The results are plotted in Fig. 11 where FMCW and SFCW show 14dB and 36dB SNR improvement over the pulsed approach, respectively. Although FMCW and SFCW should produce identical SNR improvements, 36dB as calculated based on the coherent detection sensitivity gain, the FMCW modulation is of low fidelity with its 500Hz sampling rate due to tone feedthrough (such as clock feedthrough). This artificially increases the apparent noise with FMCW which may be reduced with techniques such as chopping. Here, alternate FMCW transmit and received waveforms are phase-shifted by 180 degrees. This is equivalent to alternate signal inversion which constructively superimpose with a subtraction while the feedthrough tone is subtracted out.

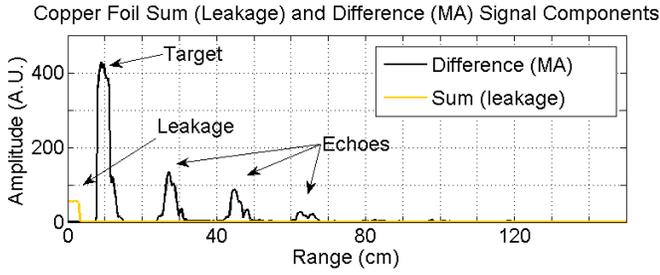

**Fig. 9**. Two measurements were taken at ±0.13T and the difference shows MA signals with echoes from the copper foil target.

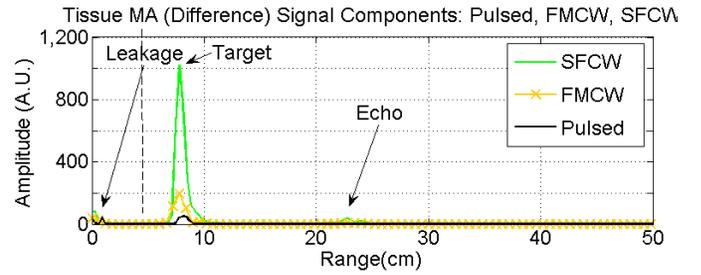

**Fig. 11**. SFCW technique shows 36dB SNR improvement over the pulsed approach with similar levels of excitation leakage at ±0.13T.

### D. Discussion

Results in Fig. 8 and Fig. 9 together confirm that the observed signals are indeed a result of the magneto-acoustic effect. RF leakage (such as EMI), parasitic ultrasound echoes (i.e. conventional US imaging), thermo-acoustic signals, electroacoustic signals, and other factors have been isolated as they do not correlate with magnetic field reversal (as in Fig. 9). The arrival time of the first ultrasound signal matches expected values corresponding to the separation of the US transducer and the target. Similarly, echoes arrive at multiples of twice this time delay corresponding to their round trip distances. The linear dependence of the US signals in Fig. 8 on both the magnetic field magnitude and RF excitation levels reaffirm that these are MA signals. Weak EA signals are present when a copper foil source is used. With small, needle-like, electrodes both EA and TA signals are produced in the tissue sample in addition to MA signals. With wider, ribbon-like electrodes, only weak EA signals in addition to MA signals are observed.

After the thorough verification of our MAI system we perform an example B-scan imaging as illustrated in Fig. 10. Here a 10mm wide copper-foil serves as a well-controlled source of MA signals under RF excitation. The resolution along the z-axis is determined by the operating bandwidth, here 250kHz. The resolution can be improved by using ultrasound transducers with higher center frequencies and bandwidths along with appropriate electronics. The resolution along the x-axis is limited to the scan step size, here 1.5mm, in addition to transducer aperture and beam widths. In the image, we clearly see the 10mm wide copper-foil centered at x=0cm and a distance of 7cm below the US transducer. There are also side-lobe artifacts visible due to the wide (13mm) beam width of the transducer, also referred to as its point spread function, which is not corrected for in the post processing. In a real imaging application, post processing in addition to the use of amplitude coded phased array transducers would be used to significantly improve the accuracy of the image by correcting for the beam pattern of the receiver.

Finally we see that the continuous wave MA techniques improve the SNR in comparison to the pulsed MA technique due to finite peak power limitation of the excitation amplifiers and the narrow pulse widths required for high resolution. It is important to note that the FMCW technique does not show as much improvement as the SFCW technique due the coarse fidelity of its implementation. An improved FMCW implementation is feasible with existing technologies, such as high speed direct digital synthesizers (e.g. AD9910).

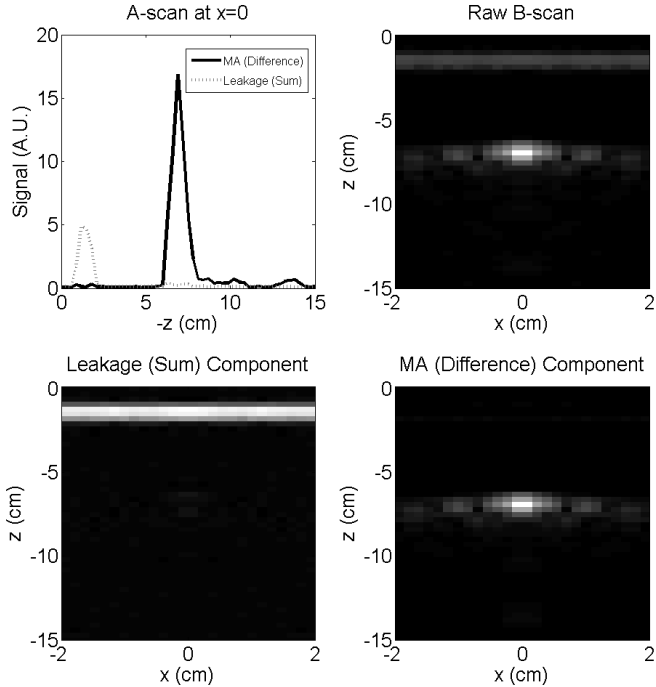

**Fig. 10**. A-scan detection (top-left) and corresponding B-scan image (top-right) of a 1-cm wide copper foil target. The leakage and MA components of the image are illustrated at the bottom-left and bottom-right respectively.

## V. Conclusion and Future Directions

The theory and detailed methodology for continuous-wave MAI was discussed. Verification was done and B-scan measurements were made in experiments deploying idealized copper-foil targets. More refined experiments demonstrated MAI with tissue samples, where continuous wave techniques reduced peak excitation powers by 36dB compared to conventional methods. This peak-power reduction can facilitate the extension of existing MAI systems by increasing their field of view to the human body size. In general, electronic components of MAI technology are amenable to silicon integration and hence miniaturization, mass-production, and cost reduction. Thus, while human-scale imaging is the main goal, the reduction of peak excitation power levels may enable new, compact, portable, and integrated MAI solutions.

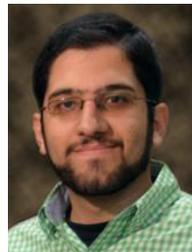

**Miaad S. Aliroteh** (S'13) received the B.A.Sc. degree (with honors) in Engineering Science with a Major in electrical & computer engineering from the University of Toronto, ON, Canada, in 2012. He is currently an electrical engineering Ph.D. candidate at Stanford University, Stanford, CA.

His research interests include multimodal biomedical imaging, biometrics, biosensing & diagnostics, lab-on-a-chip, neural interfaces, neuroprostheses, wireless implantable or wearable biomedical devices, and Analog and RF VLSI.

Mr. Aliroteh was awarded the PGS M scholarship from the Natural Sciences and Engineering Research Council of Canada in 2012 and the Qualcomm Innovation Fellowship in 2014.

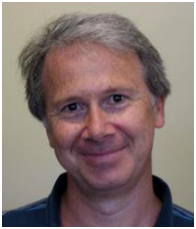

**Greg C. Scott** (M'09) received the B.A.Sc. degree (with honors) from the University of Waterloo, Waterloo, ON, Canada, in 1986, and the M.A.Sc. and Ph.D. degrees from the University of Toronto, Toronto, ON, Canada, in 1989 and 1993, respectively, all in electrical engineering.

His main research interests are magnetic resonance imaging (MRI) instrumentation and electromagnetic imaging techniques for RF safety and MR-guided therapy.

He is a Senior Research Engineer with the Magnetic Resonance Systems Research Laboratory (MRSRL), Stanford University, Stanford, CA, and has served as a consultant to several interventional device companies.

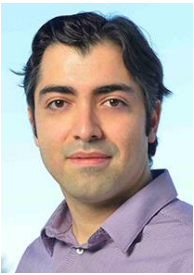

**Amin Arbabian** (S'06, M'12) received his Ph.D. degree in electrical engineering & computer science from UC Berkeley in 2011. In 2012 he joined Stanford University, as an Assistant Professor of Electrical Engineering, where he is also a School of Engineering Frederick E. Terman Fellow. In 2007 and 2008, he was part of the initial engineering team at Tagarray, Inc. He spent summer 2010 at Qualcomm's Corporate R&D division designing circuits for next generation ultra-low power wireless transceivers.

His research interests are in high-frequency circuits, systems, and antennas, medical imaging, and ultra-low power sensors. He currently serves on the TPC for the European Solid-State Circuits Conference and the Radio-Frequency Integrated Circuits (RFIC) Symposium.

Prof. Arbabian is the recipient/co-recipient of the 2015 NSF CAREER award, 2014 DARPA Young Faculty Award (YFA), 2013 IEEE International Conference on Ultra-Wideband (ICUWB) best paper award, 2013 Hellman Faculty Scholarship, 2010 IEEE Jack Kilby Award for Outstanding Student Paper at the International Solid-State Circuits Conference, two time second place Best Student Paper Awards at 2008 and 2011 RFIC symposiums, the 2009 CITRIS (Center for Information Technology Research in the Interest of Society at UC Berkeley) Big Ideas Challenge Award and the UC Berkeley Bears Breaking Boundaries award, and the 2010-11 as well as 2014-15 Qualcomm Innovation fellowships.